\documentclass[fleqn,usenatbib]{rasti}

\usepackage{newtxtext,newtxmath}
\usepackage{placeins}
\usepackage[T1]{fontenc}

\DeclareRobustCommand{\VAN}[3]{#2}
\let\VANthebibliography\thebibliography
\def\thebibliography{\DeclareRobustCommand{\VAN}[3]{##3}\VANthebibliography}

\usepackage{graphicx}	
\usepackage{amsmath}	

\title[XMST II: Galactic OB associations]{XMST II: The spatial and kinematic structure of Galactic OB associations in Gaia DR3}

\author[M. J. Gallaway]{
Mark J. Gallaway $^{1}$\thanks{E-mail: m.j.gallaway.astro@gmail.com}
\\
$^{1}$Independent Researcher, Hockley, Essex, UK
}

\date{Accepted XXX. Received YYY; in original form ZZZ}

\pubyear{\the\year{}}

\begin{document}
\label{firstpage}
\pagerange{\pageref{firstpage}--\pageref{lastpage}}
\maketitle

\begin{abstract}
Stellar associations are difficult clustering targets because position, kinematics and extinction carry different physical units, different uncertainties and different discriminatory power. We extend the Extended Minimum Spanning Tree (XMST) framework without forcing these observables into a single metric. Three-dimensional spatial connectivity first defines candidate parent structures (XMST-S1), transverse kinematics then tests those parents for dynamically distinct populations (XMST-S3), and visual extinction provides a final reddening refinement (XMST-S2). Across 1000 Monte Carlo realisations, XMST-S1 produced a better purity--completeness balance than the eight predefined three-dimensional HDBSCAN configurations tested, although neither method reliably separated a deliberately superimposed pair using spatial information alone. Applying kinematics before reddening increased mean Jaccard similarity to 0.8354 and resolved the overlapping pair in 98.6 per cent of realisations. The S3 null experiment produced only five accepted subdivisions among 106,265 tested groups. Across 1000 paired reddening-uncertainty realisations, increasing uncertainty reduced the mean number of accepted S2 subdivisions from 4.224 to 1.982, while mean Jaccard similarity changed only from 0.8074 to 0.7993. Direct XMST-S3 uncertainty tests showed negligible practical degradation for transverse-velocity errors up to $0.5~\mathrm{km\,s^{-1}}$ per component, with larger errors progressively suppressing kinematic refinement. In a label-blind repartition of 2,551 stars from 56 published OB associations, median best-match Jaccard similarity increased from 0.6535 after spatial clustering to 0.8861 after complete refinement, with 22 associations recovered identically. The main result is that physically different observables can refine an existing spatial hierarchy without redefining its metric.
\end{abstract}




\section{Introduction}

Finding stellar clusters and associations in large astrometric datasets is not simply a matter of choosing a clustering algorithm. Minimum Spanning Trees (MSTs; \citealt{Cartwright2004}), Friends-of-Friends methods \citep{Liu2019}, DBSCAN \citep{Ester1996} and HDBSCAN \citep{Campello2015} have all been used successfully, and the precision and scale of \textit{Gaia} have encouraged methods that operate simultaneously in position, parallax and proper-motion space \citep[e.g.][]{Kounkel2019,Hunt2021}. The difficulty is that these observables do not describe the same physical quantity. Combining them may improve discrimination, but it also requires a choice about how parsecs, angles and velocities should be weighted against one another.

In Paper I \citep{Gallaway2026XMST}, we introduced the Extended Minimum Spanning Tree (XMST). XMST combines percolation analysis with Jenks Natural Breaks optimisation so that the MST fracture scale is determined from the edge-length distribution itself rather than imposed as an external linking or pruning length. We validated the method with 1000 Monte Carlo realisations based on the population and spatial properties of the Gaia-derived candidate OB-star catalogue of \citet{Quintana2025}, hereafter Q25. XMST recovered the injected structures with very high completeness and cleaner membership than the alternative global fracture prescriptions tested on the same fields.

Paper I also exposed the obvious limit of a purely spatial method. If two physically distinct populations occupy nearly the same region of $(X,Y,Z)$ space, the MST can connect them into one structure. This is fundamentally a data problem, not something that can be fixed by making the spatial classifier more elaborate. If the populations are not separable in the coordinates supplied to the algorithm, additional information is needed.

Our first attempt to add reddening directly to the MST made that point rather clearly. Treating $A_V$ as another coordinate requires an arbitrary scale factor between magnitudes and parsecs, and we could not identify a physically defensible choice. We therefore changed the question. Instead of asking reddening to help define the spatial structure, we used it to test whether an already recovered spatial parent contains more than one population.

That idea is developed here. XMST-S2 uses visual extinction, $A_V$, to test for statistically distinct reddening populations, while XMST-S3 applies the same staged logic to transverse velocities in $(V_l,V_b)$. In both cases a statistically supported subdivision is accepted only if the proposed descendants remain spatially coherent when reconstructed with the original Stage-1 fracture scale. The additional observable can therefore divide an existing structure, but it cannot redefine the spatial metric that created the parent.

This staged design lets each dataset answer a specific physical question. Stage 1 asks which stars are spatially connected. The later stages ask whether a recovered parent contains populations that are distinguishable in kinematics or reddening. Because those tests are conditional, their order can matter. We therefore compare both sequences,

\[
(X,Y,Z)\rightarrow A_V\rightarrow(V_l,V_b)
\]

and

\[
(X,Y,Z)\rightarrow(V_l,V_b)\rightarrow A_V,
\]

using paired Monte Carlo fields with identical spatial and simulated velocity data. The S2 and S3 acceptance criteria are unchanged between the two runs. The preferred order is chosen from that controlled comparison before being applied to the external Q26 association catalogue.

The validation is deliberately split into separate questions. XMST-S1 is compared directly with HDBSCAN using identical three-dimensional inputs. XMST-S2 is tested on the empirical-template Monte Carlo suite. XMST-S3 is tested with a controlled velocity-separation experiment and a null experiment that measures false kinematic subdivision. We then vary the frozen evidence thresholds, add observational uncertainty to reddening and transverse velocity, and measure the effect of refinement order.

Finally, the preferred hierarchy is applied blindly to the 2,551 stars assigned to the 56 accepted Galactic OB associations of \citet{Quintana2026}, hereafter Q26. The Q26 labels are withheld during clustering and used only afterwards to measure correspondence. This is not an attempt to reproduce the Q26 algorithm. It asks whether a very different, staged clustering procedure recovers much of the same association structure.

The aim is therefore quite specific: to test whether additional astrophysical information can refine XMST in a controlled and interpretable way without sacrificing the physical meaning of its spatial metric.

\section{Method}

The spatial part of the analysis is the XMST-S1 method introduced and validated in Paper I \citep{Gallaway2026XMST}; it is not changed here. Candidate structures are identified from their three-dimensional positions using the Percolation--Jenks XMST procedure. Only after that spatial partition has been established are additional observables allowed to test the internal structure of a recovered parent.

Two refinement algorithms are used. XMST-S2 tests visual extinction, $A_V$, while XMST-S3 tests transverse velocities in $(V_l,V_b)$. The stage numbers identify the algorithms, not a fixed execution order. We test the two possible orders explicitly in Section~\ref{sec:order_test}.

The controlled order experiment favours

\[
(X,Y,Z) \rightarrow (V_l,V_b) \rightarrow A_V ,
\]

so the principal hierarchy used in this paper is XMST-S1 followed by XMST-S3 and then XMST-S2.

The S2 and S3 acceptance criteria are frozen throughout the validation experiments. Neither reddening nor velocity is folded into the spatial distance metric, and every spatial reconstruction during refinement uses the Stage-1 fracture scale. In other words, there is no conversion factor between parsecs, magnitudes of extinction and kilometres per second.

\begin{figure*}
\centering
\includegraphics[width=0.78\textwidth]{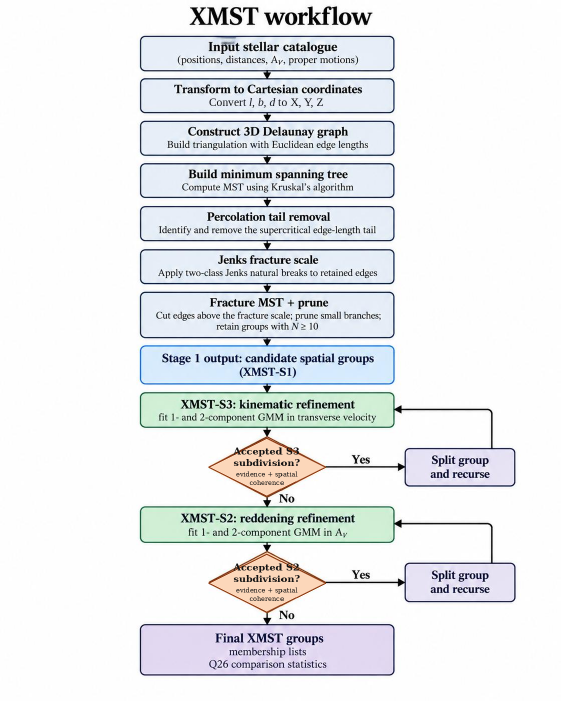}
\caption{Overview of the XMST workflow used in Paper II. The spatial
XMST-S1 catalogue is constructed using the methodology introduced in
Paper I. Candidate structures are then refined sequentially using
transverse kinematics (XMST-S3) followed by reddening (XMST-S2).
Additional observables are used only to test existing spatial structures
and are not combined with position within a common multidimensional metric.}
\label{fig:paper2_workflow}
\end{figure*}

\subsection{Input catalogue and spatial coordinates}

The parent stellar sample was taken from the catalogue of candidate O- and
B-type stars compiled by \citet{Quintana2025}, hereafter Q25. The catalogue
contains 24,706 candidate OB stars within approximately 1~kpc of the Sun.

For the spatial analysis each star is represented by its heliocentric
Galactic Cartesian coordinates $X$, $Y$ and $Z$, expressed in parsecs.
These are equivalent to

\begin{align}
X &= d \cos b \cos l,\\
Y &= d \cos b \sin l,\\
Z &= d \sin b,
\end{align}

where $d$ is heliocentric distance and $l$ and $b$ are Galactic longitude
and latitude.

No standardisation, whitening or relative rescaling of the three spatial
coordinates is applied. Euclidean distance in physical three-dimensional
space is therefore the spatial metric throughout the analysis.

The effect of the Q25 distance uncertainties on this spatial representation was examined explicitly in Paper I. Stellar distances were resampled using the published asymmetric uncertainty intervals while the sky direction of each source was held fixed, thereby introducing predominantly line-of-sight positional perturbations. For each perturbation the Cartesian coordinates, Delaunay graph, minimum spanning tree and Percolation--Jenks fracture scale were recalculated. Across 10,000 such reconstructions the global fracture scale remained stable, although the detailed membership of individual structures was less persistent, particularly for peripheral and locally sparse members. The present work therefore retains the spatial definition validated in Paper I and does not repeat the distance-uncertainty analysis.

\subsection{Stage 1: spatial XMST}

The initial spatial catalogue is generated using the XMST methodology
described in Paper I. Briefly, a three-dimensional Delaunay triangulation
is constructed from the Cartesian stellar positions and used as a sparse
candidate-edge set from which the exact Euclidean minimum spanning tree is
recovered using Kruskal's algorithm.

The global fracture scale is determined using the Percolation--Jenks
procedure. Percolation analysis first identifies the upper edge-length
regime associated with the transition from locally connected structure
towards a field-spanning network. Two-class Jenks Natural Breaks
optimisation is then applied to the remaining subcritical edge-length
distribution, and the class boundary is adopted as the fracture scale.

For the Q25 parent catalogue this gives

\[
L_{\rm fracture}=17.048~{\rm pc}.
\]

Edges longer than this value are removed and connected components containing
fewer than

\[
N_{\min}=10
\]

stars are discarded. The resulting Stage-1 catalogue contains 103 retained
groups comprising 2,339 stars.

The spatial algorithm and minimum membership criterion are held fixed for
the subsequent analysis. In particular, the Stage-1 fracture scale is not
re-estimated inside XMST-S2 or XMST-S3 descendants. Where a candidate
population is tested for spatial coherence at a later stage, a new spatial
MST is constructed from that subset of stars but is fractured using the
existing Stage-1 value of $L_{\rm fracture}$.

\subsection{Monte Carlo validation data}

The controlled validation uses the same deterministic set of 1000 Monte
Carlo realisations employed in Paper I. Each realisation contains eight
empirical stellar templates whose internal three-dimensional morphologies
were derived from structures in the Q25 catalogue.

The templates contain 192, 183, 109, 57, 51, 43, 37 and 25 stars,
respectively, giving a total injected population of 697 stars. The template
stars are removed from their original positions, leaving a fixed empirical
field containing 24,009 stars. For each realisation the templates are
rotated and repositioned before being reinserted into the field, returning
the catalogue size to 24,706 stars.

The two largest templates, T01 and T02, are deliberately positioned
approximately along the same line of sight, with radial centre separations
between approximately 57 and 83~pc. This produces a controlled case in
which two populations are difficult to distinguish from three-dimensional
spatial information alone.

The same reconstructed fields, injected memberships and random seeds are
used wherever direct comparisons between clustering configurations are
required.

\subsection{Recovery statistics}

Because the injected memberships are known, recovered structures can be
compared directly with the Monte Carlo truth. For an injected template $T$
and recovered group $G$, the number of correctly recovered stars is

\[
N_{\rm correct}=|T\cap G|.
\]

Completeness is defined as

\begin{equation}
C =
\frac{N_{\rm correct}}
     {N_{\rm injected}},
\label{eq:completeness}
\end{equation}

where $N_{\rm injected}=|T|$. Purity is

\begin{equation}
P =
\frac{N_{\rm correct}}
     {N_{\rm recovered}},
\label{eq:purity}
\end{equation}

where $N_{\rm recovered}=|G|$ includes both injected members and
contaminating stars.

Membership similarity is additionally measured using the Jaccard Index,

\begin{equation}
J =
\frac{|T\cap G|}
     {|T\cup G|}
=
\frac{N_{\rm correct}}
     {N_{\rm injected}+N_{\rm recovered}-N_{\rm correct}}.
\label{eq:jaccard}
\end{equation}

For each injected template, the recovered group containing the greatest
number of its members is selected as its best match.

An injected structure is considered detected where its best-matching group
contains at least

\[
\max\left(10,\left\lceil0.5|T|\right\rceil\right)
\]

of its members. A stricter recovery criterion requires

\[
C\geq0.80
\qquad {\rm and} \qquad
P\geq0.50.
\]

These are scoring criteria only and do not form part of the XMST-S2 or
XMST-S3 subdivision gates.

The deliberately overlapping T01--T02 pair is assessed separately. Formal
resolution requires both templates to satisfy the adopted detection
criterion and to select different recovered groups as their best matches.
Because the empirical templates may themselves contain internal structure,
we also record whether T01 and T02 occupy different best-matching
descendants even where one descendant subsequently fails the formal
completeness criterion.

\subsection{HDBSCAN benchmark}

HDBSCAN is compared with spatial XMST-S1 using the same 1000 Monte Carlo
realisations. HDBSCAN is applied directly to the raw $X$, $Y$ and $Z$
coordinates in parsecs using Euclidean distance, with no standardisation.

The minimum cluster size is fixed at

\[
\texttt{min\_cluster\_size}=10
\]

to match the minimum retained XMST membership. Four values of

\[
\texttt{min\_samples}=5,\;10,\;15,\;20
\]

are examined. Each is tested using both the excess-of-mass (EOM) and Leaf
cluster-selection methods, producing eight predefined configurations. The
calculations use

\[
\texttt{allow\_single\_cluster=False}.
\]

The complete parameter grid was specified as a sensitivity experiment and
all eight configurations are retained. No HDBSCAN configuration is selected
or modified after inspection of the recovery results.

\subsection{XMST-S2: reddening refinement}
\label{sec:s2_method}

XMST-S2 uses reddening as conditional evidence that an existing spatial
structure may contain more than one population. Reddening is not
incorporated into the spatial distance metric.

Here, a \emph{gate} denotes a predefined set of acceptance criteria that a
candidate subdivision must satisfy before proceeding to the next processing
step. The criteria are fixed by the method, while the quantities evaluated
against them are determined from the data.

For every parent structure eligible for testing, one-component and
two-component Gaussian mixture models are fitted to the distribution of
visual extinction $A_V$.

The Bayesian Information Criterion is defined as

\begin{equation}
{\rm BIC}
=
-2\ln \hat{\mathcal{L}}
+
k\ln n,
\label{eq:bic}
\end{equation}

where $\hat{\mathcal{L}}$ is the maximum likelihood of the fitted model,
$k$ is the number of fitted parameters and $n$ is the number of data points.

The preference for the two-component model is quantified as

\begin{equation}
\Delta{\rm BIC}
=
{\rm BIC}_{1{\rm G}}
-
{\rm BIC}_{2{\rm G}},
\label{eq:delta_bic}
\end{equation}

so that positive values favour the two-component model. The frozen
XMST-S2 criterion requires

\[
\Delta{\rm BIC}\geq10.
\]

The separation of the two fitted one-dimensional Gaussian components is
measured using the Ashman separation statistic \citep{Ashman1994},

\begin{equation}
D =
\frac{\sqrt{2}\,|\mu_1-\mu_2|}
     {\sqrt{\sigma_1^2+\sigma_2^2}},
\label{eq:ashman_d}
\end{equation}

where $\mu_1$ and $\mu_2$ are the fitted component means and $\sigma_1$
and $\sigma_2$ are their standard deviations. The frozen separation
criterion is

\[
D\geq2.
\]

A parent passes the XMST-S2 statistical evidence gate only where all of the
following conditions are satisfied:

\begin{enumerate}
    \item $\Delta{\rm BIC}\geq10$;
    \item Ashman $D\geq2$;
    \item each fitted Gaussian mixture model (GMM) component contains at least five stars;
    \item each fitted component contains at least 15 per cent of the 
          $A_V$ sample; and
    \item  $A_V$ measurements are available for all members of the
          tested parent.
\end{enumerate}

The final condition corresponds to a reddening-coverage requirement of
\[
f_{A_V}=1.0,
\]
where \(f_{A_V}\) is the fraction of members with  \(A_V\) measurements.

For the Q25 parent catalogue, $A_V$ measurements are available for all 24,706 stars, so this coverage requirement does not exclude any parent from XMST-S2.

It is a data-availability requirement rather than evidence for multiple
populations.

The Gaussian-mixture fits use 20 initialisations and a fixed random state
of 20260825. These settings are retained for reproducibility but are not
additional scientific acceptance thresholds.

A covariance regularisation term of $10^{-6}$ is applied to the
Gaussian-mixture fits to maintain numerical stability.

Passing the mixture-model gate does not itself determine the division
between the two populations. Once the gate has been satisfied, exact
two-class Jenks Natural Breaks optimisation is applied to the $A_V$
distribution to determine a deterministic one-dimensional dividing
threshold. Jenks therefore determines the location of the subdivision
after statistical acceptance; it is not itself the significance test.

The stars below and above the resulting $A_V$ threshold are treated as
candidate reddening populations. Each side must contain at least

\[
N_{\min}=10
\]

stars before spatial reconstruction. A new spatial MST is then independently
constructed for each population and fractured at the frozen Stage-1 spatial
scale.

Every resulting spatially coherent descendant containing at least ten stars
is retained. The spatial-child policy is therefore \texttt{all}: all
qualifying spatial descendants are preserved rather than selecting only the
largest.

For an XMST-S2 split to be accepted, both reddening populations must produce
at least one retained spatial descendant. Stars not belonging to a retained
$N\geq10$ descendant after spatial reconstruction are recorded as spatial
orphans rather than being forced into a child group.

Accepted descendants are themselves eligible for the same XMST-S2 test.
The procedure therefore operates recursively until a complete pass produces
no additional accepted split. A maximum of 20 passes is imposed as a
computational safeguard.

\subsection{XMST-S3: transverse-kinematic refinement}
\label{sec:s3_method}

XMST-S3 applies the same staged philosophy using transverse stellar
velocities. The velocity space is two dimensional and consists of
$(V_l,V_b)$; radial velocity is not required.

The transverse velocities are derived from the \textit{Gaia} proper motions
using the Q25 distances and corrected using the adopted solar-motion
convention,

\[
(U,V,W)
=
(11.1,\;12.24,\;7.25)
~{\rm km\,s^{-1}}.
\]

The resulting velocities remain expressed in the local Galactic
tangent-plane components $(V_l,V_b)$. No additional correction is
applied for perspective expansion or contraction, differential Galactic
rotation, or for transformation to a common three-dimensional bulk-motion
frame.

For every eligible parent, one-component and two-component full-covariance
Gaussian mixture models are fitted in $(V_l,V_b)$. Preference for the
two-component model uses the same definition of $\Delta{\rm BIC}$ as
Equation~\ref{eq:delta_bic}, with the frozen requirement

\[
\Delta{\rm BIC}\geq10.
\]

Because the kinematic discriminator is two dimensional, component separation
is measured using a pooled-covariance Mahalanobis distance, which quantifies
the separation of the fitted centroids relative to the joint dispersion and
covariance of the two velocity distributions. If the fitted component
centroids are $\boldsymbol{\mu}_1$ and $\boldsymbol{\mu}_2$, with covariance
matrices $\mathbf{C}_1$ and $\mathbf{C}_2$, then

\begin{equation}
D_{\rm 2D}
=
\left[
(\boldsymbol{\mu}_1-\boldsymbol{\mu}_2)^{\rm T}
\left(
\frac{\mathbf{C}_1+\mathbf{C}_2}{2}
\right)^{-1}
(\boldsymbol{\mu}_1-\boldsymbol{\mu}_2)
\right]^{1/2}.
\label{eq:d2d}
\end{equation}

The frozen separation criterion is

\[
D_{\rm 2D}\geq2.
\]

In one dimension this definition reduces to the Ashman separation statistic
used for XMST-S2.

A parent passes the XMST-S3 statistical evidence gate only where all of the
following conditions are satisfied:

\begin{enumerate}
    \item $\Delta{\rm BIC}\geq10$;
    \item $D_{\rm 2D}\geq2$;
    \item each fitted GMM component contains at least five stars;
    \item each component contains at least 15 per cent of the 
          kinematic sample; and
    \item  $(V_l,V_b)$ measurements are available for all members
          of the tested parent.
\end{enumerate}

The final condition corresponds to the kinematic-coverage requirement

\[
f_{\rm kin}=1.0.
\]

In the Q25 catalogue, 24,581 of 24,706 stars have transverse-velocity pairs. Following XMST-S2, the catalogue contains 108 retained parent groups, of which 99 satisfy the 100-per-cent kinematic-coverage requirement; the remaining nine contain at least one member without transverse kinematics and are therefore not eligible for S3 refinement.

The Gaussian-mixture fits again use 20 initialisations, a fixed
random state of 20260825 and the same covariance regularisation.

Unlike XMST-S2, no Jenks threshold is required following acceptance
of the kinematic gate. This asymmetry is deliberate: $A_V$ is a
one-dimensional ordered quantity, so once bimodality has been established
a deterministic scalar threshold can be defined. In the two-dimensional
$(V_l,V_b)$ plane there is no equivalent unique one-dimensional boundary,
and membership of the two candidate velocity populations is therefore
taken directly from the fitted two-component GMM.

Each GMM component must contain at least ten stars before spatial
reconstruction. A new spatial MST is constructed independently for each
component and fractured using the frozen Stage-1 value of
$L_{\rm fracture}$.

All resulting spatially coherent descendants containing at least ten stars
are retained. A proposed XMST-S3 division is rejected if either kinematic
component produces no retained spatial descendant.

Stars not surviving the spatial reconstruction are recorded as spatial
orphans. Accepted descendants may themselves be examined recursively using
the same frozen XMST-S3 gate, subject to a maximum of 20 passes.

No position--velocity scaling factor is introduced at any stage. Velocity
determines whether an existing spatial structure contains statistically
distinguishable populations; spatial coherence continues to be defined
solely in parsecs by the Stage-1 XMST fracture scale.

\subsection{Order-dependence experiment}
\label{sec:order_test}

Because XMST-S2 and XMST-S3 are conditional refinement operations, their
order is not mathematically guaranteed to commute. A controlled paired
experiment was therefore performed to determine whether the final hierarchy
depends materially on which additional information layer is introduced
first.

For each of the 1000 empirical-template Monte Carlo fields, two pipelines
were applied:

\[
{\rm S1}\rightarrow A_V\rightarrow(V_l,V_b)
\]

and

\[
{\rm S1}\rightarrow(V_l,V_b)\rightarrow A_V.
\]

The two pipelines used the identical reconstructed spatial field, identical
truth memberships and the same realisation of the simulated transverse
velocity field. No XMST-S2 or XMST-S3 acceptance criterion was changed
between the two orderings.

For this experiment the T01 and T02 populations were assigned
two-dimensional Gaussian transverse-velocity distributions with an
intrinsic dispersion

\[
\sigma_v=1~{\rm km\,s^{-1}}
\]

in each velocity component. Their velocity centroids were separated by

\[
\Delta V=5~{\rm km\,s^{-1}}
\]

along a randomly oriented direction in the $(V_l,V_b)$ plane. The same
velocity realisation and direction were used for both pipeline orders within
each Monte Carlo field.

The resulting recovery statistics were used to determine the principal
execution order adopted for the subsequent analysis.

\subsection{Targeted XMST-S3 velocity-separation experiment}
\label{sec:s3_velocity_test}

A separate targeted experiment was used to determine the velocity-separation
regime in which the frozen XMST-S3 discriminator begins to distinguish the
deliberately overlapping T01--T02 pair.

For this diagnostic, XMST-S3 was applied directly to the relevant Stage-1
parent without first applying XMST-S2. This isolates the information
contributed by transverse kinematics and avoids the case in which a prior
reddening subdivision has already removed the planted spatial merger.

The two injected populations were assigned Gaussian transverse velocities
with

\[
\sigma_v=1~{\rm km\,s^{-1}}
\]

per component, and centroid separations of

\[
\Delta V=
0,\;1,\;2,\;3,\;5,\;10~{\rm km\,s^{-1}}
\]

were tested. The direction of the velocity offset was randomised in the
two-dimensional velocity plane. The frozen XMST-S3 gate was unchanged at
all velocity separations.

\subsection{XMST-S3 null experiment}

The false-split behaviour of XMST-S3 was assessed using a separate set of
1000 null realisations. In these fields no distinct kinematic populations
were injected. Instead, all stars were assigned transverse velocities drawn
from a common Gaussian distribution.

The complete statistical and spatial-coherence gates were then applied
unchanged. This experiment measures the frequency with which random
velocity structure can pass both the two-component GMM evidence test and
the subsequent requirement for spatially coherent descendants.

\subsection{Sensitivity to the continuous evidence thresholds}
\label{sec:gate_sensitivity}

The numerical values adopted for the continuous S2 and S3 evidence
thresholds were subjected to a controlled sensitivity analysis.

For XMST-S2 the nominal criteria

\[
\Delta{\rm BIC}=10
\qquad\hbox{and}\qquad
D=2
\]

were multiplied simultaneously by factors

\[
0.8,\;0.9,\;1.0,\;1.1,\;1.2,
\]

corresponding to threshold pairs

\[
(8,1.6),\;
(9,1.8),\;
(10,2.0),\;
(11,2.2),\;
(12,2.4).
\]

The S2 sensitivity test was evaluated across the complete 1000-field Monte
Carlo set. For this diagnostic the S2 gate was varied in the
reddening-first sequence
\[
{\rm S1}\rightarrow {\rm S2}\rightarrow {\rm S3},
\]
with the following S3 gate held at its nominal values and the T01--T02
velocity separation fixed at $\Delta V=5~{\rm km\,s^{-1}}$. The quoted
global recovery statistics therefore describe the final partition after
the unchanged S3 stage, while the accepted-split counts refer directly to
the varied S2 stage. This is an end-to-end sensitivity test of the S2
evidence gate and is distinct from the preferred execution order established
by the paired order experiment.

The same factors were applied to $\Delta{\rm BIC}$ and $D_{\rm 2D}$ for
XMST-S3. Because the response of the kinematic stage is most directly
measured before any reddening subdivision, the S3 sensitivity test used the
targeted Stage-1-to-XMST-S3 experiment with
$\Delta V=3~{\rm km\,s^{-1}}$ and
$\sigma_v=1~{\rm km\,s^{-1}}$. For each Monte Carlo field the intrinsic
velocity realisation, offset direction and targeted T01--T02 Stage-1 parent
were identical to those used for the $\Delta V=3~{\rm km\,s^{-1}}$
condition in Table~\ref{tab:s3_velocity}; only the two continuous S3 evidence
thresholds were varied.

Throughout these sensitivity tests the component-size threshold of five
stars, component fraction of 15 per cent, data-coverage requirements,
retained child size of ten stars, \texttt{all} spatial-child policy and
fixed Stage-1 fracture scale were unchanged. The experiment therefore tests
sensitivity to the statistical evidence thresholds without altering the
structural definition of the algorithm.

\subsection{Observational-uncertainty tests}

The response of the refinement stages to observational uncertainty was
tested without altering the underlying spatial realisations, the Stage-1
fracture scale or any of the frozen acceptance criteria. The purpose of
these experiments was to determine whether increasing uncertainty in the
additional observables causes unstable subdivision, or instead progressively
reduces the evidence available for further refinement. Both the XMST-S2
reddening-uncertainty test and the targeted Stage-1-to-XMST-S3
velocity-uncertainty test were evaluated across the complete 1000-field
Monte Carlo set.

For XMST-S2, independent Gaussian perturbations were added to the
$A_V$ measurements in the empirical-template Monte Carlo fields. The Q25
catalogue does not provide a homogeneous per-star uncertainty for $A_V$,
so this was treated as a controlled absolute-noise experiment rather than
as a propagation of catalogue-derived extinction uncertainties. In addition
to the unperturbed condition, Gaussian dispersions of
\[
\sigma_{A_V} =
0.025,\ 0.05,\ 0.10,\ 0.20~{\rm mag}
\]
were examined. Within each Monte Carlo realisation, one standard-normal
reddening-error draw was reused and rescaled at each uncertainty amplitude,
so the comparison between conditions is paired. The spatial coordinates,
Stage-1 fracture scale and all XMST-S2 statistical and spatial-coherence
criteria were held fixed at every uncertainty level.

The targeted kinematic uncertainty test was performed using the
Stage-1-to-XMST-S3 experiment described in Section~\ref{sec:s3_velocity_test} and all 1000
reconstructed Monte Carlo fields. XMST-S2 was deliberately bypassed so that
the response of the S3 discriminator could be measured directly, without the
result being conditioned by any prior reddening subdivision. The intrinsic
transverse-velocity distributions retained
\[
\sigma_v = 1~{\rm km\,s^{-1}}
\]
per component and a centroid separation of
\[
\Delta V = 5~{\rm km\,s^{-1}}.
\]
For each Monte Carlo field, the intrinsic velocity realisation and the
direction of the T01--T02 velocity offset were identical to those used for
the $\Delta V=5~{\rm km\,s^{-1}}$ condition in Section~\ref{sec:s3_velocity_test}. The zero-error
condition therefore reproduces the corresponding targeted S3 result exactly.
Independent Gaussian observational perturbations were then added to both
$V_l$ and $V_b$, with per-component dispersions of
\[
\sigma_{v,{\rm obs}} =
0.25,\ 0.50,\ 1.0,\ 2.0~{\rm km\,s^{-1}}.
\]
The frozen XMST-S3 statistical and spatial-coherence gates were applied
unchanged at every uncertainty level. Within each Monte Carlo realisation,
the same standard-normal velocity-error draw was reused and rescaled across
the tested amplitudes, so these comparisons are also paired.

No spatial positions were perturbed in either experiment. The tests
therefore isolate uncertainty introduced through the additional
astrophysical observables while leaving the underlying spatial XMST
solution unchanged. They test whether degraded reddening or transverse
kinematic information propagates through the refinement hierarchy, or
whether the staged architecture instead tends to suppress subdivisions for
which the additional evidence has become insufficient.

\subsection{Blind repartition of the Q26 association population}
\label{sec:q26_method}

The final staged framework was compared with the 56 accepted Galactic OB
associations reported by \citet{Quintana2026}, hereafter Q26.

To separate the problem of member selection from the problem of partitioning
those members, the complete set of 2,551 stars assigned by Q26 to its
56 accepted associations was reanalysed. The Q26 association labels were
withheld during clustering and were used only after the XMST hierarchy had
been constructed.

Because this restricted parent sample has a substantially different
edge-length distribution from the complete 24,706-star Q25 catalogue, the
Stage-1 Percolation--Jenks scale was inferred independently from the
restricted sample using the same frozen algorithm. It was not forced to the
17.048-pc Q25 value. The resulting fracture scale should therefore be interpreted as the
algorithmic scale inferred for this deliberately restricted point process,
rather than as an alternative estimate of a characteristic physical scale
for Galactic OB associations.

The resulting spatial groups were then refined using the preferred sequence

\[
(X,Y,Z)\rightarrow(V_l,V_b)\rightarrow A_V
\]

with the unchanged XMST-S3 and XMST-S2 acceptance criteria.

Best-match completeness, purity and Jaccard similarity were calculated
against the withheld Q26 labels only after clustering was complete.
One-to-many and many-to-one mappings were retained rather than forcing a
one-to-one correspondence.

This experiment tests whether XMST partitions the same preselected stellar
population in a similar way to Q26. It is therefore a comparison of
partitioning methods applied to the same set of stars, rather than an
independent recovery of the Q26 associations or their original member
selection.

\section{Results}

The results are presented in stages. We first establish the purely spatial
baseline by comparing XMST-S1 with HDBSCAN using identical three-dimensional
inputs. We then examine the reddening and transverse-kinematic refinement
algorithms, followed by the paired order experiment, the sensitivity of the
frozen evidence gates and the response to observational uncertainty. Finally,
the preferred hierarchy is applied blindly to the stellar population selected
by Q26.

Throughout the Monte Carlo analyses, performance is measured against known
injected memberships. Comparisons with Q26 are instead catalogue-correspondence
tests and do not assume that either partition represents a unique astrophysical
ground truth.

\subsection{Spatial XMST-S1 and HDBSCAN comparison}

Across the 1000 empirical-template Monte Carlo realisations, XMST-S1
achieved a mean completeness of 0.9996 and a mean purity of 0.7413.
The mean Jaccard similarity was 0.7410, and a mean of 6.080 of the eight
injected structures satisfied the strict recovery criterion.

The performance of HDBSCAN varied substantially with both
$\texttt{min\_samples}$ and the adopted cluster-selection method.
Table~\ref{tab:hdbscan_mc} summarises the comparison.

\begin{table}
\centering
\caption{Mean recovery performance of XMST-S1 and the eight predefined
HDBSCAN configurations across 1000 Monte Carlo realisations. Strict
recovery requires $C\geq0.80$ and $P\geq0.50$.}
\label{tab:hdbscan_mc}
\begin{tabular}{lccc}
\hline
Method & $C$ & $P$ & Strict / 8 \\
\hline
XMST-S1                 & 0.9996 & 0.7413 & 6.080 \\
HDBSCAN EOM, 5          & 0.9885 & 0.5975 & 5.105 \\
HDBSCAN EOM, 10         & 0.9796 & 0.5758 & 4.921 \\
HDBSCAN EOM, 15         & 0.9966 & 0.1400 & 1.155 \\
HDBSCAN EOM, 20         & 0.9812 & 0.4808 & 4.084 \\
HDBSCAN Leaf, 5         & 0.6819 & 0.8067 & 4.309 \\
HDBSCAN Leaf, 10        & 0.7423 & 0.7889 & 4.659 \\
HDBSCAN Leaf, 15        & 0.7933 & 0.7548 & 5.126 \\
HDBSCAN Leaf, 20        & 0.8004 & 0.7313 & 5.025 \\
\hline
\end{tabular}
\end{table}

\begin{figure}
\centering
\includegraphics[width=\columnwidth]{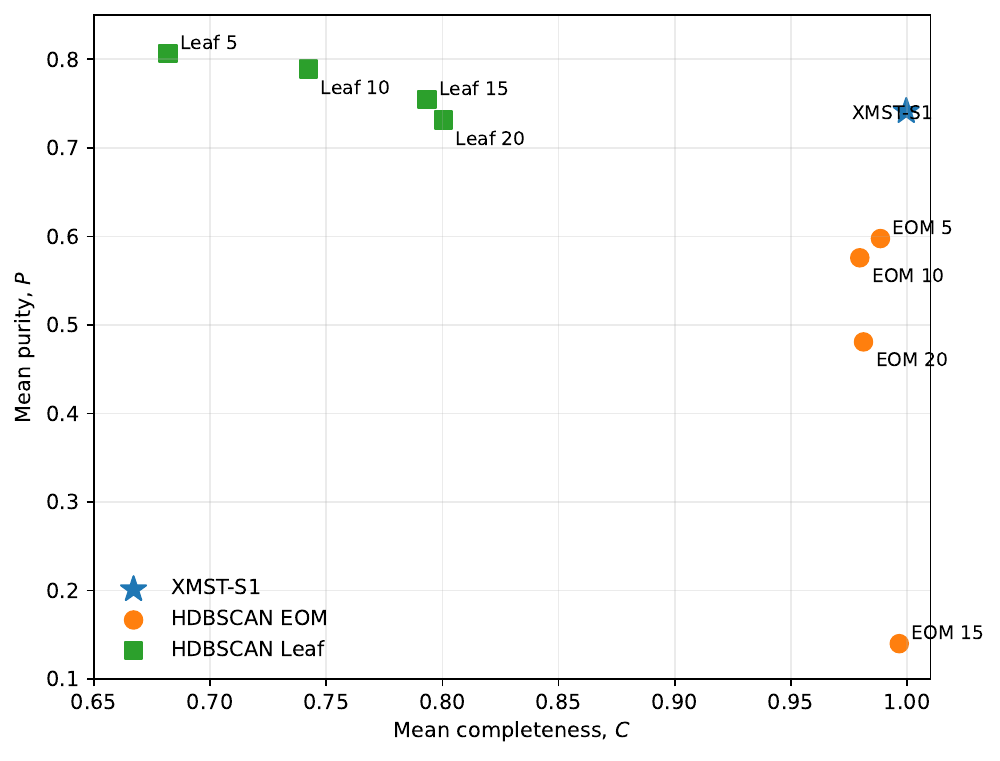}
\caption{Mean completeness and purity for XMST-S1 and the eight predefined
HDBSCAN configurations across the 1000 empirical-template Monte Carlo
realisations. XMST-S1 occupies the high-completeness, moderate-to-high-purity
region, while the HDBSCAN configurations show a stronger trade-off between
the two quantities. The labels identify the adopted \texttt{min\_samples}
values for the EOM and Leaf configurations.}
\label{fig:hdbscan_cp}
\end{figure}

The highest mean strict-recovery rate obtained by HDBSCAN was 5.126 of
eight for Leaf selection with $\texttt{min\_samples}=15$, compared with
6.080 for XMST-S1. This configuration produced only a small increase in
purity relative to XMST-S1, from 0.7413 to 0.7548, while reducing mean
completeness from 0.9996 to 0.7933.

Conversely, the high-completeness EOM configurations generally did so at the
cost of substantially increased contamination. No configuration in the
predefined HDBSCAN grid reproduced the particular combination of very high
completeness and moderate-to-high purity obtained with XMST-S1.

The deliberately superimposed T01--T02 pair remained difficult for both
purely spatial methods. XMST-S1 separated the two populations in only
15 of the 1000 realisations. HDBSCAN achieved formal separation in nine
realisations, all using EOM selection with
$\texttt{min\_samples}=5$. None of the remaining seven HDBSCAN
configurations resolved the pair under the adopted criterion.

For EOM with $\texttt{min\_samples}=5$, T01 and T02 individually retained
mean completeness values of 0.967 and 0.980, but their mean purities were
only 0.411 and 0.398. In 974 of the 1000 fields the two injected
populations selected the same HDBSCAN group as their best match.

The overlap experiment therefore demonstrates an information limit rather
than a failure specific to XMST. Where distinct populations occupy
substantially the same region of $(X,Y,Z)$ space, spatial information alone
is generally insufficient to recover them as separate structures.

\subsection{XMST-S2 reddening refinement}

XMST-S2 was evaluated using the same 1000 empirical-template realisations.
The additional reddening information changed the purity--completeness balance
substantially.

Mean completeness decreased from 0.9996 at Stage 1 to 0.9114 following
XMST-S2, while mean purity increased from 0.7413 to 0.8880. Mean Jaccard
similarity increased from 0.7410 to 0.8074 and the mean number of strict
recoveries increased from 6.080 to 6.731 of eight.

\begin{table}
\centering
\caption{Mean recovery statistics for XMST-S1 and XMST-S2 across the
1000 empirical-template Monte Carlo realisations.}
\label{tab:s2_mc}
\begin{tabular}{lcccc}
\hline
Method & $C$ & $P$ & $J$ & Strict / 8 \\
\hline
XMST-S1 & 0.9996 & 0.7413 & 0.7410 & 6.080 \\
XMST-S2 & 0.9114 & 0.8880 & 0.8074 & 6.731 \\
\hline
\end{tabular}
\end{table}

XMST-S2 increased mean purity in 99.3 per cent of the fields and increased
Jaccard similarity in 97.5 per cent. The number of strict recoveries
increased in 67.0 per cent of the realisations, was unchanged in
28.5 per cent and decreased in 4.5 per cent.

The formal overlap-resolution statistic gives an incomplete picture of the
behaviour of S2. XMST-S1 formally resolves T01 and T02 in 1.5 per cent of
the realisations, while XMST-S2 increases this only to approximately
2 per cent. In contrast, following S2 the two injected templates occupy
different best-matching descendants in 99.5 per cent of the realisations.

This difference arises primarily because T01 is itself internally
structured. The template is derived from the empirical G081 population and
is treated as one indivisible truth object in the Monte Carlo scoring. When
tested independently it strongly satisfies the frozen S2 reddening gate,
with

\[
\Delta{\rm BIC}=101.66
\]

and

\[
D=2.879.
\]

Its fitted GMM components contain 91 and 101 stars. XMST-S2 can therefore
remove the T01--T02 merger and subsequently subdivide T01 itself.
Individual descendants may then contain fewer than 50 per cent of the
original T01 membership and fail the formal truth-based overlap criterion
despite the original merger having been removed.

The reduction in nominal completeness following S2 therefore cannot be
interpreted entirely as loss of astrophysical structure. The empirical
templates reproduce internal structure from real stellar populations and
are not guaranteed to represent indivisible physical objects.

\subsection{XMST-S3 kinematic refinement}

The incremental discriminatory power of XMST-S3 was measured using the
targeted Stage-1-to-S3 experiment described in
Section~\ref{sec:s3_velocity_test}.

Little response was observed where the imposed velocity separation was

\[
\Delta V\leq2~{\rm km\,s^{-1}}.
\]

The formal resolution rate remained at the Stage-1 value of 1.5 per cent
and the global recovery statistics were effectively unchanged.

At

\[
\Delta V=3~{\rm km\,s^{-1}},
\]

the resolution rate increased to 31.0 per cent. Mean completeness remained
high at 0.9933, while purity increased to 0.7767 and mean Jaccard similarity
to 0.7707. The mean number of strict recoveries increased to 6.575 of
eight.

At

\[
\Delta V=5~{\rm km\,s^{-1}},
\]

the deliberately overlapping pair was resolved in all 1000 realisations.
Mean completeness was 0.9973, mean purity 0.8632 and mean Jaccard similarity
0.8608, with a mean of 7.932 strict recoveries.

Increasing the separation to $10~{\rm km\,s^{-1}}$ produced no further
material improvement.

\begin{table}
\centering
\caption{Targeted XMST-S3 velocity-separation experiment. T01 and T02
have intrinsic transverse-velocity dispersions of
$\sigma_v=1~{\rm km\,s^{-1}}$ per component.}
\label{tab:s3_velocity}
\begin{tabular}{rccccc}
\hline
$\Delta V$ & Resolution fraction & $C$ & $P$ & $J$ & Strict / 8 \\
\hline
0  & 0.015 & 0.9996 & 0.7413 & 0.7410 & 6.080 \\
1  & 0.015 & 0.9996 & 0.7413 & 0.7410 & 6.080 \\
2  & 0.015 & 0.9996 & 0.7413 & 0.7410 & 6.080 \\
3  & 0.310 & 0.9933 & 0.7767 & 0.7707 & 6.575 \\
5  & 1.000 & 0.9973 & 0.8632 & 0.8608 & 7.932 \\
10 & 1.000 & 0.9989 & 0.8679 & 0.8669 & 7.931 \\
\hline
\end{tabular}
\end{table}

\begin{figure}
\centering
\includegraphics[width=\columnwidth]{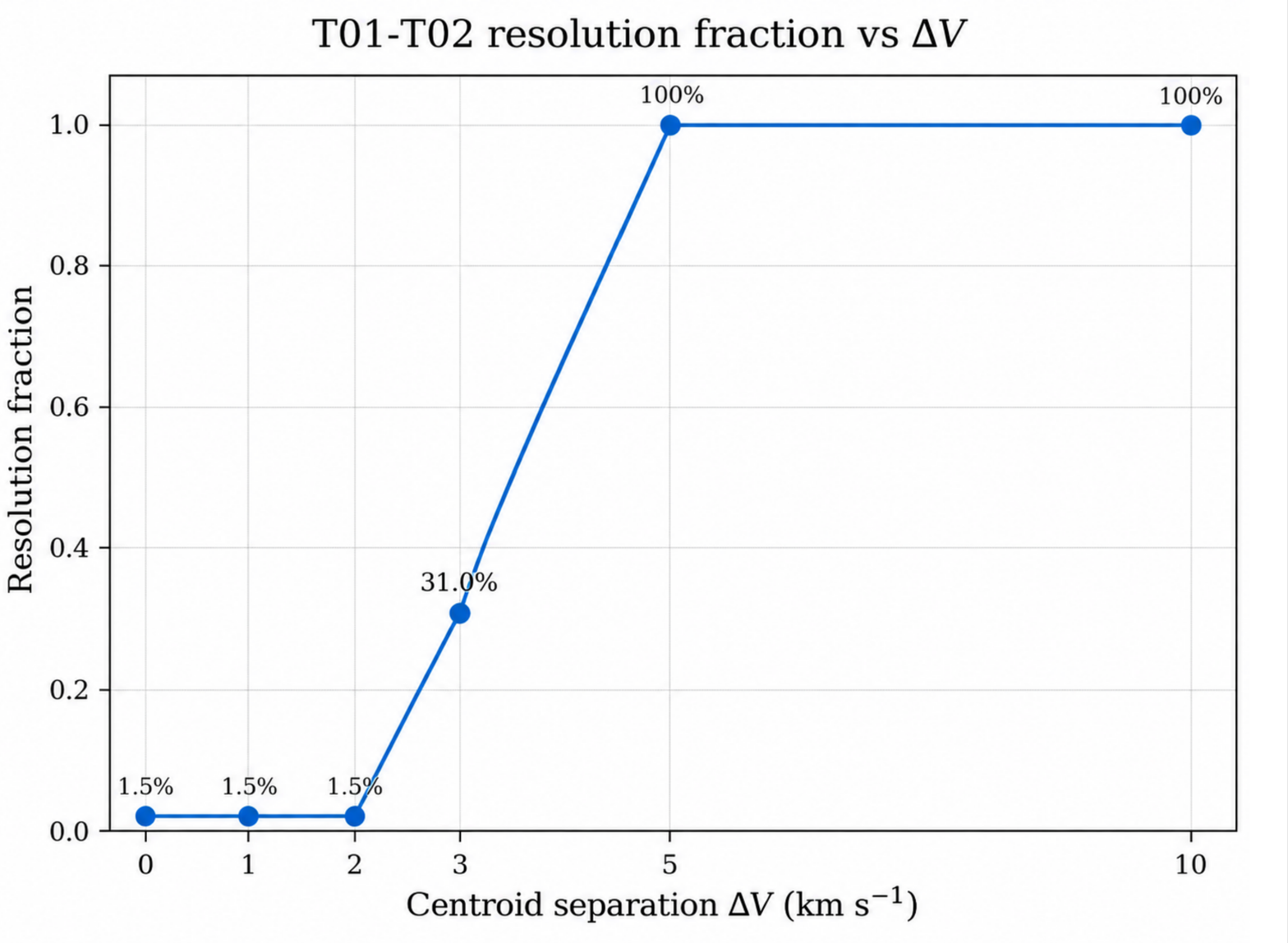}
\caption{Resolution fraction of the deliberately overlapping T01--T02
population pair as a function of the imposed transverse-velocity centroid
separation. The resolution rate remains at the Stage-1 value of 1.5 per cent
for $\Delta V \leq 2~{\rm km\,s^{-1}}$, rises to 31 per cent at
$\Delta V = 3~{\rm km\,s^{-1}}$, and reaches 100 per cent at
$\Delta V = 5~{\rm km\,s^{-1}}$.}
\label{fig:s3_resolution}
\end{figure}

At $\Delta V=5~{\rm km\,s^{-1}}$, both overlapping populations remained
almost complete rather than one simply being rejected. T01 was recovered
with $C=0.9895$, $P=0.9342$ and $J=0.9248$, while T02 had
$C=0.9918$, $P=0.9465$ and $J=0.9390$.

The null experiment provides a complementary measure of S3 behaviour. Across
1000 null fields, 106,265 active groups were examined. A total of 388
groups, approximately 0.365 per cent, passed the initial statistical GMM
criterion, but only five also satisfied the subsequent spatial-coherence
requirements and were accepted as S3 subdivisions.

Under the adopted Gaussian null model, the accepted-split rate was
therefore approximately 0.0047 per cent of all tested groups,
corresponding to accepted S3 subdivisions in 0.5 per cent of the fields. Four of the five accepted null subdivisions produced no
change in the truth-based recovery statistics. Only one affected an
injected structure and caused a single loss from the strict-recovery score.

The spatial reconstruction therefore acts as a strong physical filter on
apparently significant kinematic structure.

\begin{figure}
\centering
\includegraphics[width=\columnwidth]{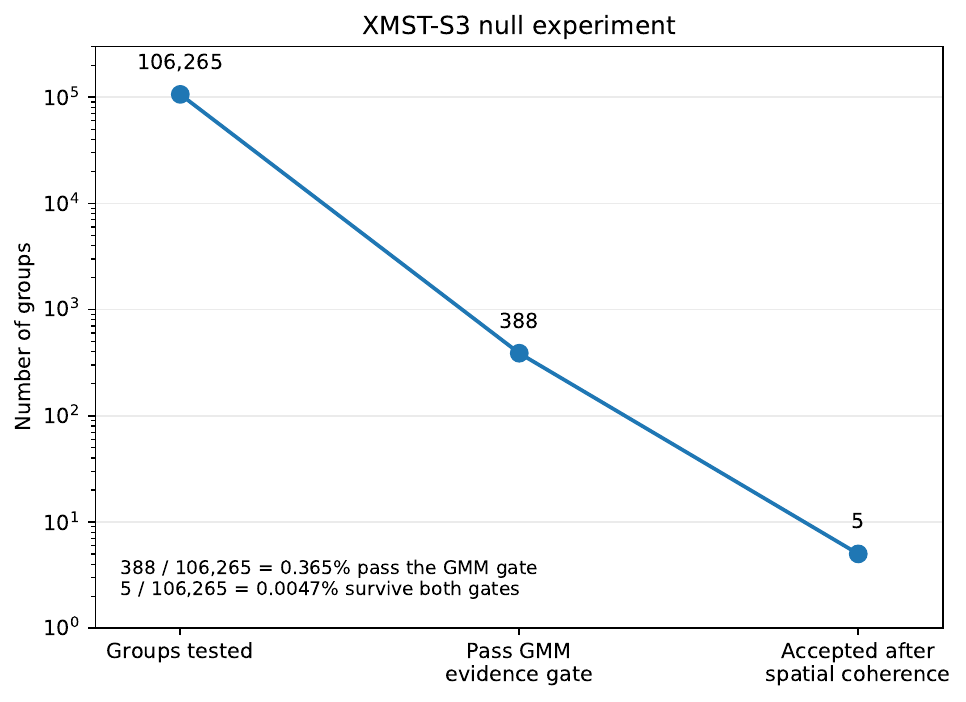}
\caption{Filtering of candidate kinematic subdivisions in the XMST-S3
null experiment. Across 1000 null realisations, 106,265 groups were tested,
of which 388 (0.365 per cent) passed the initial two-component GMM evidence
criterion. Only five groups (0.0047 per cent of those tested) subsequently
satisfied the spatial-coherence requirements and were accepted as XMST-S3
subdivisions.}
\label{fig:s3_null}
\end{figure}

\subsection{Dependence on refinement order}
\label{sec:order_results}

The paired order experiment demonstrates that XMST-S2 and XMST-S3 do not
commute. Applying the same two frozen refinement algorithms in a different
sequence can materially alter the recovered hierarchy in complex or
overlapping structures.

Both orderings were applied to the same 1000 reconstructed Monte Carlo
fields using identical velocity realisations:

\[
{\rm S1}\rightarrow A_V\rightarrow(V_l,V_b)
\]

and

\[
{\rm S1}\rightarrow(V_l,V_b)\rightarrow A_V.
\]

For this experiment the T01--T02 pair had
$\Delta V=5~{\rm km\,s^{-1}}$ and
$\sigma_v=1~{\rm km\,s^{-1}}$ per velocity component.

The reddening-first ordering produced a mean completeness of approximately
0.9113, mean purity of 0.8887 and mean Jaccard similarity of 0.8079,
with 6.741 strict recoveries of eight. The formal overlap-resolution rate
was 2.5 per cent.

Reversing the two refinement stages produced a mean completeness of 0.9420,
mean purity of 0.8884 and mean Jaccard similarity of 0.8354. The mean
number of strict recoveries increased to 7.000 of eight.

Most notably, the formal T01--T02 overlap-resolution rate increased to
98.6 per cent. The paired mean change in Jaccard similarity was
$\Delta J=+0.02750$ (95 per cent bootstrap CI $[0.02594,0.02908]$).
The corresponding increase in the formal T01--T02 overlap-resolution
fraction was 0.961 (95 per cent bootstrap CI $[0.948,0.973]$).

\begin{table}
\centering
\caption{Paired comparison of the two refinement orders across 1000
identical Monte Carlo fields.}
\label{tab:order}
\begin{tabular}{lcccc}
\hline
Order & $C$ & $P$ & $J$ & Strict / 8 \\
\hline
S1$\rightarrow$S2$\rightarrow$S3
 & 0.9113 & 0.8887 & 0.8079 & 6.741 \\
S1$\rightarrow$S3$\rightarrow$S2
 & 0.9420 & 0.8884 & 0.8354 & 7.000 \\
\hline
\end{tabular}
\end{table}

\begin{figure*}
\centering
\includegraphics[width=\textwidth,trim=0 75bp 0 0,clip]{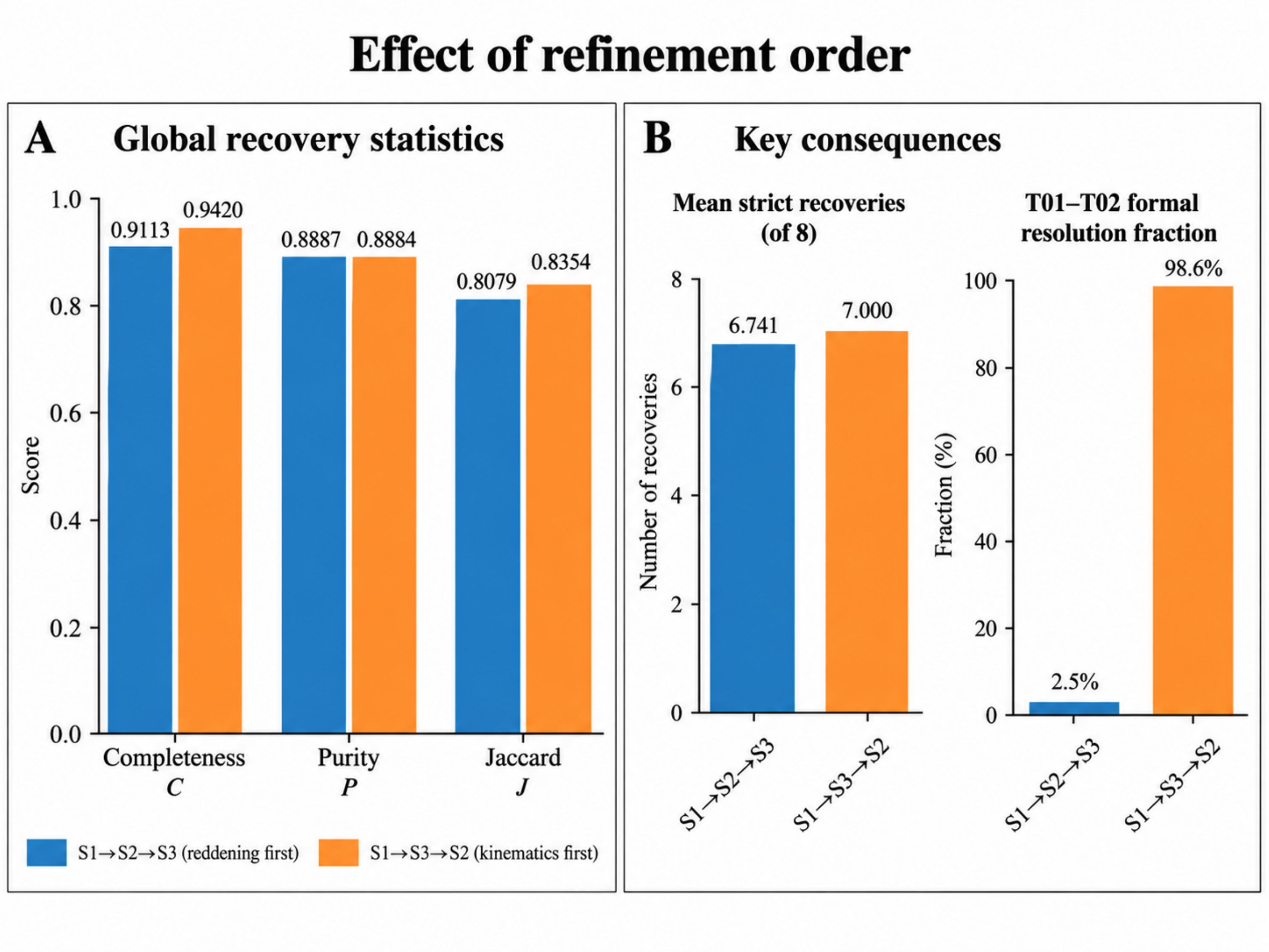}
\caption{Effect of refinement order on the recovered hierarchy. Panel A
compares the global recovery statistics for the two refinement sequences.
Panel B shows the corresponding mean strict-recovery count and the formal
T01--T02 overlap-resolution fraction. Applying kinematics before reddening,
$(X,Y,Z)\rightarrow(V_l,V_b)\rightarrow A_V$, improves completeness,
Jaccard similarity and strict recovery, while purity remains essentially
unchanged. The most striking difference is the formal T01--T02 resolution
fraction, which increases from 2.5 per cent for the reddening-first ordering
to 98.6 per cent for the kinematics-first ordering.}
\label{fig:refinement_order}
\end{figure*}

The change in purity is negligible, whereas completeness and Jaccard
similarity both increase. The difference is therefore not simply produced
by accepting more contaminants.

The preferred hierarchy for the remainder of this work is consequently

\[
(X,Y,Z)\rightarrow(V_l,V_b)\rightarrow A_V.
\]

Kinematics first separates populations that can be distinguished
dynamically before the reddening algorithm examines the resulting
descendants for additional internal structure.

\FloatBarrier

\subsection{Sensitivity to the continuous evidence thresholds}
\label{sec:gate_results}

The continuous S2 and S3 evidence thresholds were varied by
$\pm20$ per cent around their adopted values while all structural criteria
were retained unchanged.

For the XMST-S2 gate test, the global recovery statistics were measured
after the following frozen S3 stage. Relaxing the S2 gate to
$\Delta{\rm BIC}=8$ and $D=1.6$ produced
$C=0.8932$, $P=0.8916$ and $J=0.7950$.
At the nominal gate, the corresponding values were
$C=0.9114$, $P=0.8879$ and $J=0.8073$.
Tightening the gate to
$\Delta{\rm BIC}=12$ and $D=2.4$ produced
$C=0.9313$, $P=0.8749$ and $J=0.8104$.

The number of accepted S2 subdivisions declined smoothly as the evidence
requirements became more restrictive, from a mean of 4.972 at the most
permissive setting to 3.088 at the most restrictive, compared with 4.224
at the nominal gate.

The targeted S3 sensitivity experiment shows a similarly smooth response.
At $\Delta V=3~{\rm km\,s^{-1}}$, relaxing the thresholds to
$\Delta{\rm BIC}=8$ and $D_{\rm 2D}=1.6$ produced a formal resolution
rate of 35.4 per cent and mean $J=0.7752$. At the nominal gate the
resolution rate was 31.0 per cent with $J=0.7707$, exactly reproducing the
corresponding condition in Table~\ref{tab:s3_velocity}. Tightening the gate
to $\Delta{\rm BIC}=12$ and $D_{\rm 2D}=2.4$ reduced the resolution rate
to 25.8 per cent while mean $J$ remained 0.7655.

\begin{table*}
\centering
\caption{Sensitivity of the continuous S2 and S3 evidence thresholds.
The factor multiplies both the nominal $\Delta{\rm BIC}=10$ threshold
and the corresponding separation threshold of 2. For the S2 rows, the
S2 gate is varied in the reddening-first S1$\rightarrow$S2$\rightarrow$S3
diagnostic and $C$, $P$ and $J$ are measured after the following frozen S3
stage. The S3 rows use the targeted Stage-1-to-S3 diagnostic at
$\Delta V=3~{\rm km\,s^{-1}}$.}
\label{tab:gate_sensitivity}
\begin{tabular}{lcccccc}
\hline
Stage & Factor & $\Delta{\rm BIC}$ & Separation & $C$ & $P$ & $J$ \\
\hline
S2 & 0.8 & 8  & 1.6 & 0.8932 & 0.8916 & 0.7950 \\
S2 & 0.9 & 9  & 1.8 & 0.8981 & 0.8905 & 0.7983 \\
S2 & 1.0 & 10 & 2.0 & 0.9114 & 0.8879 & 0.8073 \\
S2 & 1.1 & 11 & 2.2 & 0.9222 & 0.8847 & 0.8115 \\
S2 & 1.2 & 12 & 2.4 & 0.9313 & 0.8749 & 0.8104 \\
\hline
S3 & 0.8 & 8  & 1.6 & 0.9923 & 0.7821 & 0.7752 \\
S3 & 0.9 & 9  & 1.8 & 0.9930 & 0.7789 & 0.7726 \\
S3 & 1.0 & 10 & 2.0 & 0.9933 & 0.7767 & 0.7707 \\
S3 & 1.1 & 11 & 2.2 & 0.9938 & 0.7737 & 0.7682 \\
S3 & 1.2 & 12 & 2.4 & 0.9944 & 0.7705 & 0.7655 \\
\hline
\end{tabular}
\end{table*}

There is no abrupt transition at the adopted gate values. Changing the statistical evidence thresholds changes how often subdivisions are accepted, but the recovery statistics move smoothly across the tested interval. The nominal S3 gate also reproduces exactly the Table~\ref{tab:s3_velocity} $\Delta V=3~{\rm km\,s^{-1}}$ result, confirming that the sensitivity experiment differs only in the adopted thresholds. The nominal result is not a narrow tuning artefact.

\begin{figure}
\centering
\includegraphics[width=\columnwidth]{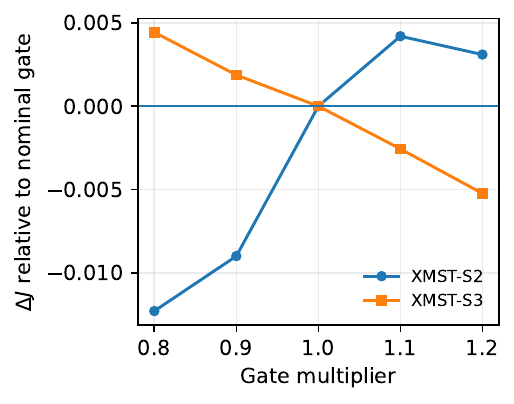}
\caption{Sensitivity of XMST-S2 and XMST-S3 to the continuous evidence
thresholds. The plotted quantity is the change in mean Jaccard similarity,
$\Delta J$, relative to the nominal gate for each diagnostic. The nominal
$\Delta{\rm BIC}$ and separation criteria are multiplied simultaneously
by factors from 0.8 to 1.2. Both diagnostics vary smoothly across the tested
interval, with no abrupt change at the adopted threshold. The S2 curve gives
the final S1$\rightarrow$S2$\rightarrow$S3 recovery score after varying
only the S2 gate; the S3 curve is from the targeted
$\Delta V=3~{\rm km\,s^{-1}}$ Stage-1-to-S3 experiment.}
\label{fig:gate_sensitivity}
\end{figure}

\FloatBarrier

\subsection{Robustness to observational uncertainty}

Both observational-uncertainty experiments were evaluated across the
complete set of 1000 paired Monte Carlo realisations. Increasing uncertainty
primarily reduced the amount of accepted refinement; in the direct S3 test
this effect becomes substantial once the added velocity uncertainty is
comparable to or larger than the adopted intrinsic dispersion.

For XMST-S2, the unperturbed realisations gave mean completeness
$C=0.9114$, mean purity $P=0.8880$ and mean Jaccard similarity
$J=0.8074$, with 6.731 strict recoveries from the eight injected
structures. At $\sigma_{A_V}=0.20$ mag these values were
$C=0.9215$, $P=0.8732$ and $J=0.7993$, with 6.835 strict recoveries.
The response in $J$ was shallow and non-monotonic: modest added reddening
noise slightly improved the truth-based score before the largest perturbation
reduced it below the unperturbed value.

Relative to the zero-error condition, the paired mean changes in Jaccard
similarity were $+0.00420$ at $\sigma_{A_V}=0.025$ mag
(95 per cent bootstrap CI $[0.00321,0.00519]$), $+0.00556$ at
$0.05$ mag ($[0.00430,0.00683]$), $+0.00252$ at $0.10$ mag
($[0.00099,0.00402]$) and $-0.00810$ at $0.20$ mag
($[-0.01015,-0.00608]$).

The clearest systematic response was a progressive reduction in the number
of accepted XMST-S2 subdivisions. The mean number of accepted splits
decreased from 4.224 in the unperturbed realisations to 3.871, 3.229,
2.326 and 1.982 for $\sigma_{A_V}=0.025$, 0.05, 0.10 and 0.20 mag,
respectively. Thus, as the reddening distributions become less
discriminating, the S2 evidence gate increasingly retains the existing
parent structure rather than accepting additional subdivisions. The modest
initial increase in $J$ is consistent with marginal subdivisions sometimes
being suppressed before the information loss becomes large enough to reduce
the overall correspondence with the injected populations.

\begin{figure}
\centering
\includegraphics[width=\columnwidth]{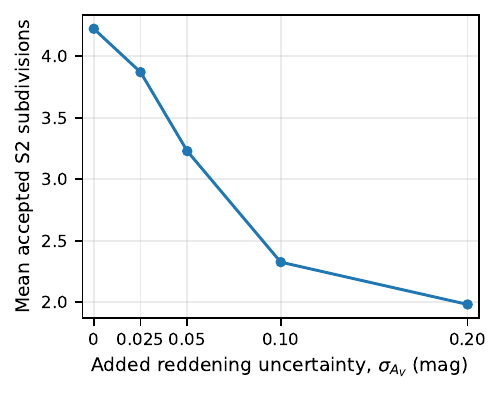}
\caption{Mean number of accepted XMST-S2 subdivisions per realisation as a
function of added reddening uncertainty across 1000 paired Monte Carlo
realisations. Increasing $\sigma_{A_V}$ produces a smooth decline in the
number of accepted splits, from 4.224 in the unperturbed case to 1.982 at
$\sigma_{A_V}=0.20$ mag. The behaviour is consistent with increasing
uncertainty progressively suppressing unsupported refinement rather than
destabilising the underlying spatial hierarchy.}
\label{fig:s2_uncertainty}
\end{figure}

A different response is obtained when XMST-S3 is measured directly,
before any reddening refinement. Across 1000 paired realisations, the
zero-error experiment gave $C=0.9973$, $P=0.8632$ and $J=0.8608$, exactly
reproducing the $\Delta V=5~\mathrm{km\,s^{-1}}$ condition in
Table~\ref{tab:s3_velocity}, with formal resolution of the T01--T02 pair in
all 1000 fields. At
$\sigma_{v,\mathrm{obs}}=0.5~\mathrm{km\,s^{-1}}$ per component,
mean $J$ decreased only to 0.8586 and the pair remained formally resolved
in 999 of 1000 realisations. At $1.0~\mathrm{km\,s^{-1}}$, mean $J$
decreased to 0.8306 and the formal overlap-resolution rate to 82.5 per cent.
At $2.0~\mathrm{km\,s^{-1}}$, mean $J$ fell to 0.7410 and the overlap-
resolution rate to 1.5 per cent, while the mean strict-recovery count
returned to 6.080 of eight. Thus, in this controlled
$\Delta V=5~\mathrm{km\,s^{-1}}$, $\sigma_v=1~\mathrm{km\,s^{-1}}$
configuration, the direct S3 discriminator is practically insensitive to
modest observational errors of up to about $0.5~\mathrm{km\,s^{-1}}$
per component, but loses discriminatory power as the observational
uncertainty approaches and exceeds the adopted intrinsic velocity
dispersion.

\begin{table}
\centering
\caption{Targeted XMST-S3 observational-uncertainty experiment across
1000 paired Monte Carlo realisations. The planted T01--T02 populations have
$\Delta V=5~\mathrm{km\,s^{-1}}$ and intrinsic per-component dispersion
$\sigma_v=1~\mathrm{km\,s^{-1}}$. The quoted velocity uncertainty is the
additional Gaussian observational dispersion per component.}
\label{tab:s3_uncertainty}
\begin{tabular}{rccccc}
\hline
$\sigma_{v,\rm obs}$ & Resolution fraction & $C$ & $P$ & $J$ & Strict / 8 \\
\hline
0.00 & 1.000 & 0.9973 & 0.8632 & 0.8608 & 7.932 \\
0.25 & 1.000 & 0.9968 & 0.8631 & 0.8604 & 7.932 \\
0.50 & 0.999 & 0.9953 & 0.8627 & 0.8586 & 7.928 \\
1.00 & 0.825 & 0.9893 & 0.8402 & 0.8306 & 7.573 \\
2.00 & 0.015 & 0.9996 & 0.7413 & 0.7410 & 6.080 \\
\hline
\end{tabular}
\end{table}

Relative to the zero-error condition, the paired mean changes in Jaccard
similarity were $-0.00047$ at
$0.25~\mathrm{km\,s^{-1}}$ (95 per cent bootstrap CI
$[-0.00059,-0.00034]$), $-0.00226$ at
$0.5~\mathrm{km\,s^{-1}}$ ($[-0.00261,-0.00198]$), $-0.03020$ at
$1.0~\mathrm{km\,s^{-1}}$ ($[-0.03288,-0.02755]$) and $-0.11985$ at
$2.0~\mathrm{km\,s^{-1}}$ ($[-0.12075,-0.11886]$).

The failure mode is conservative. As the additional observables become less informative, XMST accepts fewer subdivisions. In the direct S3 experiment, sufficiently large kinematic uncertainties drive the recovery statistics back towards the Stage-1-only solution rather than towards a new and unstable partition. Uncertainty suppresses refinement; it does not reorganise the underlying spatial hierarchy.

\subsection{Blind repartition of the Q26 association population}
\label{sec:q26_results}

The complete set of 2,551 stars assigned to the 56 accepted Q26
associations was reanalysed with the published association labels withheld
during clustering.

The spatial XMST-S1 stage produced 36 groups containing all 2,551 stars.
The Percolation--Jenks fracture scale inferred from this restricted parent
population was 61.675~pc, substantially larger than the 17.048-pc value
obtained from the complete Q25 catalogue. This difference reflects the
different edge-length distribution of the preselected association sample
rather than a change in the fracture-scale methodology.

The median best-match Jaccard similarity between the Stage-1 partition and
the withheld Q26 labels was 0.6535.

Applying XMST-S3 kinematic refinement first increased the number of groups
from 36 to 48, with 2,536 stars retained. The median best-match Jaccard
similarity increased to 0.7973.

At this stage, 44 of the 56 Q26 associations had a best-matching XMST
counterpart with $J\geq0.5$, 28 had $J\geq0.8$, 26 had $J\geq0.9$,
and 23 were recovered with identical membership.

Subsequent XMST-S2 reddening refinement produced 58 groups containing
2,526 stars. The median best-match Jaccard similarity increased further
to 0.8861.

In the final kinematics--reddening catalogue, 47 of the 56 Q26 associations
have a best-matching XMST group with $J\geq0.5$, 30 have $J\geq0.8$,
26 have $J\geq0.9$, and 22 are recovered with identical membership.

\begin{table*}
\centering
\caption{Blind XMST repartition of the 2,551 stars selected as members
of the 56 Q26 associations. Q26 association labels were withheld during
clustering.}
\label{tab:q26_order}
\begin{tabular}{lrrrrrrr}
\hline
Stage & Groups & Stars & Median $J$ &
$J\geq0.5$ & $J\geq0.8$ & $J\geq0.9$ & $J=1$ \\
\hline
XMST-S1 spatial
& 36 & 2551 & 0.6535 & 33 & 25 & 22 & 22 \\
XMST-S3 kinematics
& 48 & 2536 & 0.7973 & 44 & 28 & 26 & 23 \\
XMST-S3$\rightarrow$XMST-S2
& 58 & 2526 & 0.8861 & 47 & 30 & 26 & 22 \\
\hline
\end{tabular}
\end{table*}

For comparison, the previous reddening--kinematics ordering produced
54 groups containing 2,531 stars and a median best-match Jaccard similarity
of 0.8868.

The near equality of the two final median values hides differences in a
minority of complex structures. Of the 56 Q26 associations, 12 improve in
best-match Jaccard when kinematics is applied first, 41 are unchanged and
three worsen. The mean per-association change in Jaccard is $+0.0332$.

The two final partitions nevertheless remain globally similar. Among the
2,518 stars grouped by both orderings, the adjusted Rand index is 0.93445
and the normalised mutual information is 0.97895. Following optimal
one-to-one matching of groups, 2,388 of those 2,518 stars, 94.84 per cent,
occupy corresponding groups.

The order dependence is therefore concentrated in a minority of complex or
blended regions rather than globally restructuring the catalogue.

Some of the largest improvements when kinematics is applied first occur in
Cyg OB10b, whose best-match Jaccard increases from 0.090 to 0.720,
Her OB1b from 0.161 to 0.750, Her OB1a from 0.447 to 0.649,
Tr.~10a from 0.510 to 0.706, and Cyg OB10a from 0.468 to 0.613.
Car OB3a improves from $J=0.882$ to an exact match. The largest reduction
occurs for Gem OB3, whose best-match Jaccard decreases from 0.679 to 0.413.

The Q26 result mirrors the controlled order test. Applying kinematics first changes relatively little of the catalogue as a whole, but it materially improves several of the most strongly blended systems.

\begin{figure}
\centering
\includegraphics[width=\columnwidth]{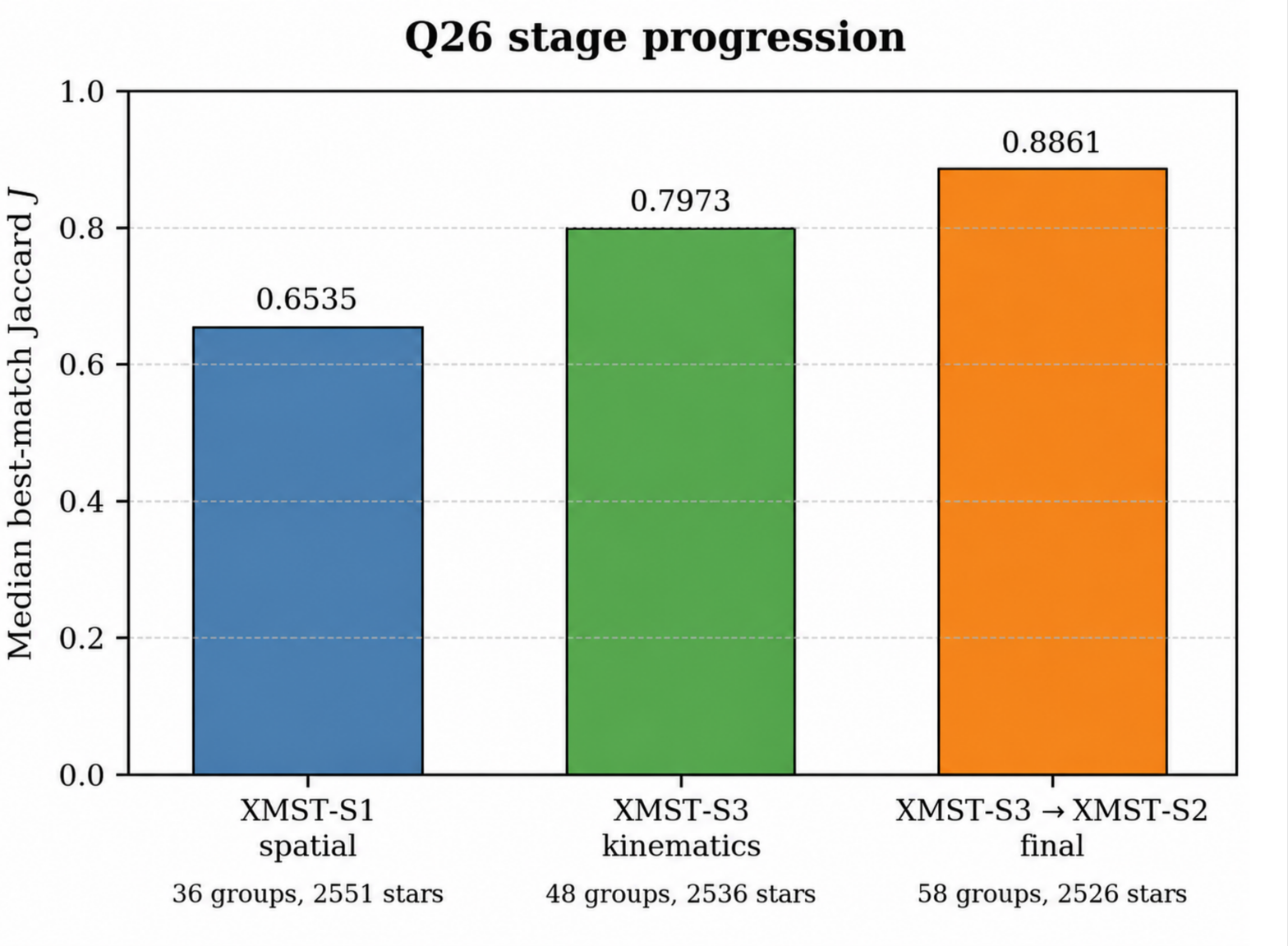}
\caption{Progression of the median best-match Jaccard similarity between
the XMST partition and the withheld Q26 association labels. Spatial
XMST-S1 gives a median similarity of $J=0.6535$, which increases to
$J=0.7973$ following XMST-S3 kinematic refinement and to $J=0.8861$
after subsequent XMST-S2 reddening refinement.}
\label{fig:q26_progression}
\end{figure}

Figure~\ref{fig:q26_progression} shows that the correspondence with the
withheld Q26 labels increases progressively as independent physical
information is introduced into the hierarchy.

\FloatBarrier

\section{Discussion}

The main advantage of staged XMST is simple: position, velocity and reddening do not have to be forced into the same metric. Stage 1 defines the spatial parent. Kinematics and extinction are then allowed to ask whether that parent contains more than one population, and any proposed subdivision still has to make sense spatially.

That distinction matters for Galactic OB associations because the available observables have different units, different uncertainties and different discriminatory power. A joint metric can certainly be useful, but it has to encode choices about their relative scaling. XMST makes a different choice. It keeps the spatial definition fixed and lets the additional observables act conditionally. The experiments here show what that choice buys us, and where it still fails.

\subsection{Spatial clustering and the information limit}

The HDBSCAN comparison establishes the spatial baseline. Across the controlled Monte Carlo fields, XMST-S1 recovered a combination of very high completeness and moderate-to-high purity that none of the eight predefined three-dimensional HDBSCAN configurations reproduced. The highest HDBSCAN strict-recovery rate was 5.126 of eight injected structures, compared with 6.080 for XMST-S1. That HDBSCAN configuration gained only a small amount of purity while reducing mean completeness from 0.9996 to 0.7933.

That is not a claim that XMST is universally better than HDBSCAN. HDBSCAN is configuration-dependent and different datasets may favour different choices. The narrower result is enough: on the same three-dimensional fields and against the same injected memberships, XMST-S1 gives the better overall purity--completeness balance among the configurations tested.

The deliberately overlapping T01--T02 pair is more interesting because both methods fail for the same reason. XMST-S1 separated the pair in only 15 of 1000 realisations, and the tested HDBSCAN configurations managed at most nine. In the best HDBSCAN case for this pair, both injected populations were largely recovered, but usually into the same group. The stars were found; the partition was not.

This is an information limit, not an algorithm-specific defect. If two physical populations occupy substantially the same $(X,Y,Z)$ volume, they cannot be expected to be separated reliably from spatial information alone. The sensible response is not automatically a more complicated spatial classifier. It is independent astrophysical information.

\subsection{Conditional refinement beyond the spatial limit}

XMST-S2 and XMST-S3 use that extra information without changing the spatial metric. Their success is the main test of the staged design.

Reddening changes the purity--completeness balance substantially. XMST-S2 increases mean purity from 0.7413 to 0.8880 and mean Jaccard similarity from 0.7410 to 0.8074, while the mean number of strict recoveries rises from 6.080 to 6.731 of eight. The fall in nominal completeness needs more care. The empirical templates come from real stellar structures and are not guaranteed to be indivisible physical populations. T01, in particular, contains strong internal reddening structure and independently passes the frozen S2 bimodality criteria. S2 can therefore remove the planted T01--T02 merger and then split T01 itself, even though the scoring treats T01 as a single truth object.

That explains the apparently awkward overlap statistic. The strict truth-based definition records only a small increase in formal T01--T02 resolution after S2, but the two injected populations occupy different best-matching descendants in 99.5 per cent of the realisations. The lesson is important: an injected membership is exact as a numerical label, but it is not guaranteed to be an astrophysically indivisible object.

The S3 velocity experiment is cleaner. With an intrinsic dispersion of $\sigma_v=1~{\rm km\,s^{-1}}$ per component, S3 shows little response at centroid separations of $0$--$2~{\rm km\,s^{-1}}$, begins to distinguish the populations near $3~{\rm km\,s^{-1}}$, and resolves the overlapping pair in every realisation at $\Delta V=5~{\rm km\,s^{-1}}$ and above. These are properties of the controlled experiment, not universal physical boundaries. The useful point is the scaling: kinematics becomes effective once the separation is large enough relative to the internal velocity dispersion.

The null experiment matters just as much. Of 106,265 tested groups, 388 passed the initial two-component GMM evidence test, but only five survived the spatial-coherence reconstruction. The accepted false-split rate is therefore about $0.0047$ per cent of all tested groups. That number belongs to the adopted Gaussian null model, not to all possible real velocity fields, but the role of the spatial reconstruction is clear. It is not a cosmetic final check. It removes most apparently significant kinematic subdivisions that do not survive as coherent spatial structures.

\subsection{Why refinement order matters}

S2 and S3 do not commute. The order matters in practice, not just mathematically.

Applying kinematics before reddening gives higher mean completeness and Jaccard similarity than applying reddening first, while purity is essentially unchanged. Most strikingly, the formal resolution rate of the deliberately overlapping T01--T02 pair rises to 98.6 per cent for
\[
(X,Y,Z)\rightarrow(V_l,V_b)\rightarrow A_V.
\]

The reason is physically straightforward. Kinematics first sees the larger spatial parent and can separate dynamically distinct populations before either is further divided by its internal reddening structure. If reddening goes first, real extinction structure within one planted population can split that population before S3 ever sees the original merger. The final catalogues remain globally similar, but the difference matters in the most strongly blended systems.

This is why the refinement stages should not be thought of as interchangeable extra dimensions. Their order is part of the physical question being asked: which stars are spatially connected; does that parent contain dynamically distinct populations; and, within those descendants, is there additional extinction structure?

The kinematics-first sequence is the preferred order for the experiments and Q26 application in this paper. I would not treat it as a universal ordering for every stellar population. The size of the effect will depend on the relative kinematic and reddening contrasts, the internal structure and the degree of spatial overlap.

\subsection{Robustness to thresholds and observational uncertainty}

A method that works only at one finely tuned threshold is not much use. That is not what we see here. Varying both $\Delta{\rm BIC}$ and the corresponding separation threshold by $\pm20$ per cent changes the frequency of accepted subdivisions and the recovery statistics smoothly. There is no sharp transition at $\Delta{\rm BIC}=10$ or separation $=2$.

The observational-uncertainty tests ask a different, and more practical, question: what happens when the extra information itself becomes worse?

For reddening, increasing the imposed uncertainty from zero to $\sigma_{A_V}=0.20$ mag reduces the mean number of accepted S2 subdivisions from 4.224 to 1.982. Over the same range, mean Jaccard similarity changes only from 0.8074 to 0.7993, a paired mean change of $-0.00810$ with a 95 per cent bootstrap CI of $[-0.01015,-0.00608]$, while the mean strict-recovery count changes from 6.731 to 6.835 of eight. At the smaller perturbations, mean $J$ rises slightly before declining at the largest uncertainty. The simplest interpretation is that a little noise suppresses some marginal, truth-disfavoured subdivisions before larger information loss starts to reduce the discriminatory value of $A_V$ itself.

That is exactly the behaviour we want from a staged method. Increasing uncertainty does not alter the spatial metric. It simply makes the extra observable less able to justify another split, so the algorithm increasingly keeps the existing spatial parent.

The direct Stage-1-to-S3 experiment shows the same idea in kinematics. Performance is effectively unchanged for observational uncertainties of $0.25$--$0.5~\mathrm{km\,s^{-1}}$ per component, begins to decline at $1~\mathrm{km\,s^{-1}}$, and is largely lost by $2~\mathrm{km\,s^{-1}}$. Combining intrinsic and observational dispersions in quadrature reduces the planted separation from $\Delta V/\sigma_{\rm eff}=5.0$ in the unperturbed experiment to 3.54 at $1~\mathrm{km\,s^{-1}}$ and 2.24 at $2~\mathrm{km\,s^{-1}}$. Again, these values describe this controlled $\Delta V=5~\mathrm{km\,s^{-1}}$, $\sigma_v=1~\mathrm{km\,s^{-1}}$ experiment, not universal S3 thresholds.

At the largest uncertainty, the recovery statistics move back towards the Stage-1-only result because S3 stops accepting subdivisions. As the extra information gets worse, XMST falls back towards the simpler spatial solution rather than inventing a new hierarchy.

This separation of roles is useful. Noise in $A_V$ or transverse velocity is largely confined to the stage that uses it; it does not redefine the spatial neighbourhoods that created the parent catalogue. The method therefore degrades conservatively: strong extra evidence can separate populations that overlap spatially, while weak evidence leaves the spatial solution largely intact.

\subsection{Correspondence with the Q26 associations}

The blind Q26 experiment is an external test of the resulting partition, but it is important to be precise about what was tested. We repartition the 2,551 stars already selected as members of the 56 Q26 associations. The Q26 labels are hidden during clustering and used only afterwards. This is a test of partition correspondence, not an independent discovery experiment.

The correspondence improves as independent information is added. Median best-match Jaccard similarity rises from 0.6535 after S1 to 0.7973 after kinematic refinement and to 0.8861 after the subsequent reddening stage. In the final hierarchy, 47 of the 56 Q26 associations have a best-matching XMST structure with $J\geq0.5$, 30 have $J\geq0.8$, 26 have $J\geq0.9$, and 22 are recovered with identical membership.

That level of agreement is striking because the clustering architectures are very different. It suggests that much of the nearby OB-association structure identified by Q26 is strong enough to reappear without reproducing the Q26 procedure itself.

The disagreements are at least as interesting as the exact matches. OB associations are hierarchical, and there is no reason to expect a continuous stellar hierarchy to possess one unique discrete partition. One method may retain a coherent subgroup separately while another includes it in a larger association; nearby substructures may equally be combined into a wider complex by a different connectivity rule.

Those discrepant systems are therefore obvious targets for independent tests. Radial velocities, stellar ages, colour--magnitude structure, gas and dust morphology and other evidence of common origin can tell us whether one partition has the stronger physical interpretation, or whether the right answer is itself hierarchical.

\subsection{Implications and limitations of the staged XMST approach}

The main methodological problem is not simply finding spatial overdensities. It is deciding when one spatially connected structure contains more than one physically meaningful population.

XMST separates those questions. Stage 1 measures connectivity in parsecs. S2 and S3 then bring in independent evidence for internal structure without inventing a conversion between parsecs, magnitudes of extinction and kilometres per second. The method still contains numerical acceptance criteria, but it avoids an additional set of scale factors specifying how unlike observables should be weighted against one another. We do not attempt a general comparison with every possible joint multidimensional clustering scheme here. The claim is narrower: the staged architecture avoids an imposed cross-observable metric and keeps the physical role of each information layer visible.

There are clear limitations. The current hierarchy uses one fracture scale derived from the relevant Stage-1 parent catalogue. That deliberately preserves the spatial definition of Paper I, but a single global scale cannot adapt to large local density changes within the same field.

S3 also uses transverse rather than full three-dimensional velocities, and S2 is only as useful as the available extinction measurements. Paper I tested the net effect of predominantly radial distance perturbations on XMST recovery, but did not classify individual membership changes separately as radial chaining or fragmentation.

The 100-per-cent coverage requirement is deliberately conservative. It simplifies the validation experiments, but it can also prevent otherwise useful tests in incomplete catalogues. In the present Q25 application it has no effect on S2 because $A_V$ coverage is complete, while 99 of the 108 S2 parent groups satisfy the corresponding kinematic requirement for S3. A parent that fails this test is retained at its current hierarchical level; it is not discarded. Future implementations will need a principled treatment of incomplete coverage rather than simply lowering the existing thresholds.

An accepted subdivision is not, by itself, proof that the descendants are distinct physical OB associations. XMST identifies statistically supported, spatially coherent structure. Age, origin and dynamical history still have to be established astrophysically.

A natural next step is to relax the assumption that the same fracture scale applies at every level of the hierarchy. A recovered parent could define a local spatial domain within which Percolation--Jenks is evaluated again. That would allow the characteristic spatial scale to respond to local structure while keeping spatial discovery separate from the additional evidence used for refinement. This local-scale extension is left for subsequent work.

XMST-S3 also needs to be interpreted for what it is: a discriminator in derived transverse-velocity space, not a complete dynamical decomposition. Perspective effects, differential Galactic rotation and changes in the local tangent basis across an extended association can all introduce coherent structure in $(V_l,V_b)$. The controlled Monte Carlo experiments validate S3 for prescribed transverse-velocity distributions; they do not validate corrections for those projection effects. An accepted S3 subdivision means kinematically distinct in the adopted observable space. It does not, on its own, establish distinct three-dimensional moving groups.

The broader point is that the stages have different jobs. XMST-S1 supplies the spatial baseline; S3 and S2 add conditional evidence only where the spatial information is insufficient. The resulting hierarchy is competitive with the tested spatial HDBSCAN configurations, reproduces much of the independently derived Q26 association structure, and tends back towards the spatial solution as the additional observables lose precision. That is a useful alternative to treating every available observable as another coordinate in one clustering space.

\section{Conclusions}

We extended the spatial XMST framework of Paper I with two conditional refinement stages: transverse kinematics (XMST-S3) and reddening (XMST-S2). The key design choice is not to put position, velocity and extinction into one combined metric. Spatial connectivity defines the parent structure first; the additional observables then test whether that parent should be subdivided, and any accepted descendants must remain spatially coherent.

The main conclusions are:

\begin{enumerate}

\item The spatial baseline is strong, but it has a real information limit. Across 1000 controlled Monte Carlo realisations, XMST-S1 achieved mean completeness of 0.9996 and mean purity of 0.7413, with 6.080 of the eight injected structures satisfying the strict recovery criterion. The best mean strict-recovery rate among the eight predefined three-dimensional HDBSCAN configurations was 5.126 of eight. More importantly, neither method could reliably separate the deliberately superimposed T01--T02 pair from spatial information alone: XMST-S1 resolved it in 15 of 1000 realisations and the tested HDBSCAN configurations in at most nine. If two populations overlap in $(X,Y,Z)$, more spatial machinery does not create information that is not there.

\item Conditional refinement adds useful information without redefining the spatial metric. XMST-S2 increased mean purity from 0.7413 to 0.8880 and mean Jaccard similarity from 0.7410 to 0.8074, with strict recoveries increasing from 6.080 to 6.731 of eight. The targeted S3 experiment showed little response below about $3~{\rm km\,s^{-1}}$ for the adopted $\sigma_v=1~{\rm km\,s^{-1}}$ field, but resolved the overlapping pair in every realisation at $\Delta V=5~{\rm km\,s^{-1}}$ and above. In the S3 null experiment only five subdivisions survived both the statistical and spatial-coherence tests from 106,265 groups.

\item The order of refinement matters. In paired experiments using the same spatial fields and velocity realisations,
\[
(X,Y,Z)\rightarrow(V_l,V_b)\rightarrow A_V
\]
produced mean completeness of 0.9420 and mean Jaccard similarity of 0.8354 with essentially unchanged purity, and resolved the deliberately overlapping pair in 98.6 per cent of realisations. This kinematics-first ordering is preferred for the experiments and Q26 application considered here; it is not asserted as a universal ordering for all stellar populations.

\item The method is not balanced on a narrow choice of thresholds and it fails conservatively as the extra information becomes worse. Varying the continuous S2 and S3 evidence thresholds by $\pm20$ per cent produced smooth changes rather than a sharp transition. Across 1000 paired reddening-uncertainty realisations, increasing $\sigma_{A_V}$ to $0.20$ mag reduced the mean number of accepted S2 subdivisions from 4.224 to 1.982 while mean Jaccard similarity changed only from 0.8074 to 0.7993. In the direct 1000-field S3 uncertainty test, $0.5~\mathrm{km\,s^{-1}}$ per-component errors produced negligible practical degradation; larger errors progressively suppressed kinematic refinement. As the extra information loses discriminatory power, XMST tends back towards the Stage-1 spatial solution rather than generating a new unstable hierarchy.

\item The blind Q26 repartition gives an external check on the staged hierarchy. With the Q26 labels withheld during clustering, median best-match Jaccard similarity rises from 0.6535 after S1 to 0.7973 after kinematic refinement and 0.8861 after subsequent reddening refinement. In the final hierarchy, 47 of the 56 Q26 associations have $J\geq0.5$, 30 have $J\geq0.8$, 26 have $J\geq0.9$, and 22 are recovered with identical membership. The remaining disagreements should not automatically be treated as errors: OB associations are hierarchical, and different physically defensible partitions may exist in complex regions.

\end{enumerate}

The central result is not simply that XMST performs competitively with the tested spatial HDBSCAN configurations. It is that physically different observables can be used to refine structure that spatial data alone cannot resolve without sacrificing the meaning of the original spatial metric. The stages keep separate questions separate.

The descendants should still be treated as candidate physical structures rather than automatically as distinct OB associations. Radial velocities, ages, colour--magnitude structure, gas and dust morphology and other independent evidence remain necessary for astrophysical interpretation, especially in the systems where XMST and Q26 disagree.

The present implementation also keeps a single Stage-1 fracture scale throughout the hierarchy. Allowing recovered parents to define local domains in which Percolation--Jenks is re-evaluated is the natural next development.

\section*{Acknowledgements}

This work has made use of data from the European Space Agency (ESA) mission Gaia (https://www.cosmos.esa.int/gaia), processed by the Gaia Data Processing and Analysis Consortium (DPAC). Funding for the DPAC has been provided by national institutions, in particular the institutions participating in the Gaia Multilateral Agreement.

The author also thanks the authors of the Quintana et al.\ catalogue for making their data available.

Generative AI tools were used during manuscript preparation for language editing and to assist with coding. All scientific decisions, analysis, code execution, verification and interpretation were carried out by the author. The analysis code was tested using an independent implementation written from the method specification rather than derived from the production code, and the two implementations were compared blindly on deterministic synthetic and Monte Carlo datasets. The spatial XMST-S1 implementation produced exact membership agreement across 100 independent 24,706-star Monte Carlo fields, and the complete XMST-S1→XMST-S3→XMST-S2 pipeline produced exact membership agreement across 10 full controlled fields. The production code was also exercised through a pytest test suite covering the principal expected behaviours and regression checks.
\section*{Data Availability}

The stellar catalogue used in this work is publicly available through the
VizieR catalogue service as catalogue J/MNRAS/538/1367
\citep{Quintana2025}.
The XMST source code, configuration files and compact validation outputs used
in this work are available from the XMST Paper II GitHub repository,
https://github.com/mjgallawayastro-stack/XMST-Paper-II. Version 1.0.0 of the
reproducibility archive, including the frozen analysis code, compact validation
outputs and manuscript source, is archived on Zenodo at DOI
10.5281/zenodo.22830657. The Q25/Q26 source catalogues are not redistributed
with the software archive and should be obtained from their original published
sources.

\section*{Funding}

The author declares no funding.

\section*{Conflict of Interest}

The author declares no conflict of interest.


\bibliographystyle{rasti}
\bibliography{PaperII} 

@article{Gallaway2026XMST,
    author  = {Gallaway, M.},
    title   = {{XMST}: An Extended Minimum Spanning Tree Framework with Objective Fracture-Scale Selection},
    journal = {RAS Techniques and Instruments},
    year    = {2026},
    note    = {arXiv:2609.12943 Submitted}
}

@article{Quintana2026,
  author   = {Quintana, A. L. and Wright, N. J. and Kormann, L. A. and Alves, J. and Katz, D. and Casamiquela, L. and Di Matteo, P. and Haywood, M. and Laporte, C.},
  title    = {Mapping the Solar Neighbourhood with {Gaia}},
  journal  = {MNRAS},
  fjournal = {Monthly Notices of the Royal Astronomical Society},
  volume   = {549},
  pages    = {stag853},
  year     = {2026}
}

@article{Cartwright2004,
    author  = {Cartwright, A. and Whitworth, A. P.},
    title   = {The statistical analysis of star clusters},
    journal = {MNRAS},
    volume  = {348},
    pages   = {589--598},
    year    = {2004},
    doi     = {10.1111/j.1365-2966.2004.07360.x}
}

@article{Liu2019,
    author  = {Liu, L. and Pang, X.},
    title   = {A Catalog of Newly Identified Star Clusters in {Gaia} {DR2}},
    journal = {ApJS},
    volume  = {245},
    pages   = {32},
    year    = {2019},
    doi     = {10.3847/1538-4365/ab530a}
}

@inproceedings{Ester1996,
    author    = {Ester, M. and Kriegel, H.-P. and Sander, J. and Xu, X.},
    title     = {A Density-Based Algorithm for Discovering Clusters in Large Spatial Databases with Noise},
    booktitle = {Proceedings of the Second International Conference on Knowledge Discovery and Data Mining},
    pages     = {226--231},
    year      = {1996}
}

@article{Campello2015,
    author  = {Campello, R. J. G. B. and Moulavi, D. and Zimek, A. and Sander, J.},
    title   = {Hierarchical Density Estimates for Data Clustering, Visualization, and Outlier Detection},
    journal = {ACM Transactions on Knowledge Discovery from Data},
    volume  = {10},
    number  = {1},
    pages   = {5:1--5:51},
    year    = {2015},
    doi     = {10.1145/2733381}
}

@article{Kounkel2019,
    author  = {Kounkel, M. and Covey, K.},
    title   = {Untangling the {Galaxy}. I. Local Structure and Star Formation History of the {Milky Way}},
    journal = {AJ},
    volume  = {158},
    pages   = {122},
    year    = {2019},
    doi     = {10.3847/1538-3881/ab339a}
}

@article{Hunt2021,
    author  = {Hunt, E. L. and Reffert, S.},
    title   = {Improving the open cluster census. I. Comparison of clustering algorithms applied to {Gaia} {DR2} data},
    journal = {A\&A},
    volume  = {646},
    pages   = {A104},
    year    = {2021},
    doi     = {10.1051/0004-6361/202039341}
}

@article{Quintana2025,
    author  = {Quintana, Alexis L. and Wright, Nicholas J. and Mart{\'i}nez Garc{\'i}a, Juan},
    title   = {A census of {OB} stars within 1 kpc and the star formation and core collapse supernova rates of the {Milky Way}},
    journal = {Monthly Notices of the Royal Astronomical Society},
    year    = {2025},
    volume  = {538},
    number  = {3},
    pages   = {1367--1383},
    doi     = {10.1093/mnras/staf083}
}

@article{Ashman1994,
  author  = {Ashman, Keith M. and Bird, Christina M. and Zepf, Stephen E.},
  title   = {Detecting Bimodality in Astronomical Datasets},
  journal = {The Astronomical Journal},
  year    = {1994},
  volume  = {108},
  number  = {6},
  pages   = {2348--2361},
  doi     = {10.1086/117248}
}





\bsp	
\label{lastpage}
\end{document}